\documentclass[showpacs,preprintnumbers,amsmath,amssymb]{revtex4}
 \usepackage{amssymb,amsmath}
 \usepackage{graphicx,epsfig}
 \usepackage{graphicx}
 \DeclareGraphicsExtensions{.pdf,.png,.jpg}
 \usepackage{dcolumn}
 \usepackage{bm}

\begin{document}

 \title{A MAGNETO-GRAVITATIONAL TRAP FOR PRECISION STUDIES OF GRAVITATIONAL QUANTUM STATES}

\author{V.V. Nesvizhevsky $^1$, F. Nez $^2$, S.A. Vasiliev $^3$, E. Widmann $^4$, P. Crivelli $^5$, S. Reynaud $^2$, A.Yu. Voronin $^6$}

\address{
$^1$ Institut Max von Laue -– Paul Langevin, 71 avenue des Martyrs, Grenoble, France, F-38042\\
$^2$ Laboratoire Kastler Brossel, Sorbonne Universit\'e, CNRS, ENS-PSL Universit\'e, Coll\`ege de France, 75252, Paris, France\\
$^3$ Department of Physics and Astronomy, University of Turku, Turku, Finland, FI-20014\\
$^4$ Stefan Meyer Institute for Subatomic Physics, Austrian Academy of Sciences, Boltzmanngasse 3, A-1090 Wien, Austria\\
$^5$ ETH, Zurich, Institute for Particle Physics and Astrophysics, 8093 Zurich, Switzerland\\
$^6$ Lebedev Institute, 53 Leninsky pr., Moscow, Russia, Ru-119333}

\begin{abstract}
Observation time is the key parameter for improving the precision of measurements of gravitational quantum states of particles levitating above a reflecting surface. We propose a new method of long confinement in such states of atoms, anti-atoms, neutrons and other particles possessing a magnetic moment. The Earth gravitational field and a reflecting mirror confine particles in the vertical direction. The magnetic field originating from electric current passing through a vertical wire confines particles in the radial direction. Under appropriate conditions, motions along these two directions are decoupled to a high degree. We estimate  characteristic parameters of the problem, and list possible systematic effects that limit storage times due to the coupling of the two motions. In the limit of low particle velocities and magnetic fields, precise control of the particle motion and long storage times in the trap can provide ideal conditions for both gravitational, optical and hyperfine spectroscopy: for the sensitive verification of the equivalence principle for antihydrogen atoms; for increasing the accuracy of optical and hyperfine spectroscopy of atoms and antiatoms; for improving constraints on extra fundamental interactions from experiments with neutrons, atoms and antiatoms.
\end{abstract}

\maketitle

\section{Introduction}

Gravitational quantum states (GQSs) of light neutral particles (hydrogen ($H$), deuterium ($D$), helium ($He$), anti-hydrogen ($\bar{H}$) atoms, neutrons ($n$) and others) levitating above a reflecting surface can be used to improve the precision of various measurements \cite{lus78,nes98,nes02,nes03,nes03reply,ber03,abel03,rob04,frank04,nes05,ber05,kie05,beber05,brau06,math06,brax07,ber06,rom07,sah07,buis07,
baes07,del09,noz10,kaj10,nes10PU,vor11,anto11,jen11,ped11,brax11,kob11,chai11,gar12,ame13,duf14,vor14,criv14,jen14,bel14,ich14,bas15,vor16,vas19,nes19,crep19}.

Observation time is the key parameter that controls the improved precision. We propose a new method of long confinement of neutral particles possessing a magnetic moment in GQSs in a Magneto-Gravitational Trap (MGT). A key feature of the new method is the combination of the vertical confinement by gravity and quantum reflection from a mirror and the radial confinement by the magnetic field of a vertical linear current. Both confinement principles are well established \cite{sch95,sch99,sch99PRL,nes02,bren15}. However, the combination of the two approaches seems to be challenging as the magnetic field might produce large false effects thus making impossible any precision studies of GQSs. We show that one could achieve small mixing of radial and vertical motions and thus can control false effects to an acceptable degree.

The most typical trap for cold atoms is the Magneto-Optical Trap (MOT) \cite{raab87}. It combines magnetic trapping and optical cooling. Such a trap was used in the ground-breaking experiments on Bose-Einstein Condensate (BEC) in a gas of ultracold atoms \cite{dav95,and95}. Since there is no maximum magnetic field in 3D, low-field-seeking (lfs) atoms are trapped in MOTs in the field minimum. Since they are not in the lowest internal state, any disturbance (collisions of atoms, magnetic field inhomogeneities) can flip their spin. This magnetic relaxation to the untrapped state is the main mechanism of losses from MOTs. However, it is absent for traps for high-field-seeking (hfs) atoms, which may allow the trapping of clouds of atoms of much higher density. Dynamic magnetic traps for hfs atoms based on rapidly varying electromagnetic fields have been proposed but they are typically shallow.

In contrast, the MGT provides a deep trapping potential and is especially suitable for the lightest alkalis: $H$ and $D$. For $H$, the trap barrier height of $\sim 0.5$ K can be easily realized, and allows trapping of a large number of atoms at temperatures of $\sim 100$ mK. Optical cooling methods based on the $1S-2P$ or two-photon $1S-2S$ transitions can be used down to the recoil limit of $\sim 2$ mK. Further cooling of the trapped gas can be done using evaporation over the trap barrier. Atomic collisions in the high-density regime provide high equilibration rate, leading to substantially lower temperatures.

In the low-density collisionless regime, the hfs particles undergo adiabatic motion along closed trajectories around the conducting wire. At low particle velocities, precise control of their motion and sufficiently large storage times in GMTs can provide ideal conditions for both gravitational, optical and hyperfine spectroscopy: for the sensitive verification of the equivalence principle for $\bar{H}$, for increasing the accuracy of optical and hyperfine spectroscopy of atoms and antiatoms, for improving constraints on extra fundamental interactions from experiments with $n$, atoms and antiatoms.

The principle of operation of the proposed MGT for precision studies of GQSs is described in section II. Coupling of vertical and radial motions of the particles in the MGT is analyzed in section III. A feasibility of loading/unloading the MGT as well as examples of precision measurements of GQSs in the MGT are presented in section IV.

We focus on the properties of the MGT but leave for later, more detailed publications the crucial topics of loading/unloading the trap and spectroscopy/ interferometry of GQSs. We will only discuss briefly these topics noting that methods of spectroscopy/interferometry of GQSs have been developed in detail \cite{nes10NP,pig14,vor14,bas15,crep19}, and that fast changes of the electric current can load/unload the MGT in case of atoms and anti-atoms, while super-fluid $^4He$ in the trap exposed to the flux of cold neutrons allows producing ultra-cold neutrons (UCNs) directly in the trap \cite{golub75}.

\section{Description of the trap}
\begin{figure}
  \centering
\includegraphics[width=100mm]{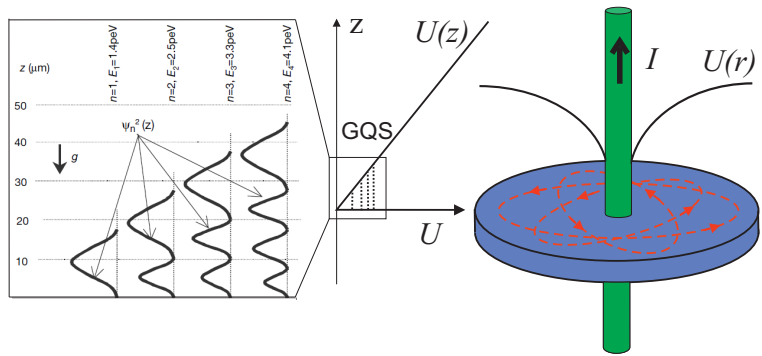}
\caption{\label{trap} On the right side, a schematic representation of the Magneto-Gravitational trap (MGT) for neutral particles with a magnetic moment is shown. The linear gravity potential and quantum reflection from the horizontal mirror (blue) confine particles in the vertical direction. The particle's wave functions in four lowest GQSs are shown in the insert on the left as a function of the height above the mirror. The magnetic field of a vertical linear wire carrying an electric current $I$ and the centrifugal acceleration confine particles in the horizontal plane. The magnetic field is inversely proportional to the distance to the wire. The particle adiabatically moves along closed elliptical trajectories (red dashed lines around the wire), similar to the orbital motion of planets around the sun.}
\end{figure}

Fig. \ref{trap} shows a scheme of the MGT. The mirror and gravitational field confine particles vertically. The interaction of a particle's magnetic moment with the vertical electric current and the centrifugal acceleration confine particles radially.

Under appropriate conditions that will be discussed in the paper, in particular for a sufficiently long wire, forces acting on particles in the vertical and horizontal directions are almost orthogonal so that the corresponding motions are decoupled to a high degree. The vertical motion of particles with lowest vertical energies is governed by quantum mechanics while the horizontal motion can be considered classical in realistic conditions.

\subsection{Motion of a magnetic dipole in the field of a linear current}
The quantum motion of a magnetic dipole in the magnetic field of linear current can be described analytically \cite{pron77,blum89}. Here, an adiabatic approximation, which allows classical treatment, is sufficient for our purpose. It relies on the hierarchy of characteristic times associated with the fast spin motion and the slowly varying magnetic field in the frame co-moving with the particle in the plane perpendicular to the current. In the MGT, the particle trajectories follow Kepler-like orbits around the wire. For the simplest case of a circular orbit with a radius $r$, the rotation frequency (in Hz) of a magnetic dipole $\mu$ is given by:
\begin {equation}\label{angular velocity}
\Omega^2=\frac{\mu_{0}\mu I}{2\pi r^3 m}=\frac{B(r)\mu}{r^2 m},
\end{equation}
where $I$ is the electric current, $\mu_{0}$ the magnetic permeability of vacuum, $m$ the particle mass, and $r$ a radial distance from a given point to the wire carrying the current. To meet the adiabaticity condition, the angular velocity of rotation has to be much smaller than the Larmor frequency of the spin precession, $\Omega\ll\omega_L=\gamma B(r)$, with $\gamma$ being the gyromagnetic ratio for the magnetic dipole under consideration. Radial confinement will occur in the Coulomb-like potential:
\begin {equation}\label{Urad}
U(r)=-\mu B(r)=-\frac{\mu_0 \mu I}{2\pi r},
\end{equation}
with the Bohr energy and the Bohr radius of the particle
\begin{equation}
E_B=\frac{m}{2\hbar^2}\left(\frac{\mu_0 \mu I}{2\pi}\right)^2, \mbox{ } R_B=\frac{2\pi \hbar^2}{m \mu_0 \mu I }.
\end{equation}
The Coulomb potential depth reaches a maximum at the wire surface. The value of the potential strength depends on the type of particle; it is roughly three orders of magnitude larger for electron spins than for nuclear spins, or for the neutron spin. Therefore, one needs much stronger magnetic fields and electric currents for trapping $n$ or $^3$He.

Let's evaluate the feasibility of achieving magnetic trapping for $n$ with their small magnetic moments. The current density in the wire is $i_0=I/{\pi R^2}$, where the wire radius $R$ is a free parameter. A typical value for the most common superconductors based on NbTi is $i_0\approx 10^5$ A/cm$^2$ in the magnetic field of $3$ T and the temperature of $4.2$ K. With this current density constraint, the potential depth as a function of the wire radius is

\begin{equation}
U(R)=\frac{1}{2} \mu_0\mu i_0 R.
\end{equation}

It looks that increasing the wire radius is beneficial for making a stronger magnetic trap. This is, however, only true until we reach another constraint associated with the maximum critical current density. Then the current density has to be reduced. This effect depends on the type and manufacture of the wire and the temperature. For the values given above and a wire radius of $0.5$ cm, the field near the wire surface is $B_{max}\sim 3$ T, which is close to the critical value. Therefore, the specified maximum field and the trap depth are reached at a wire radius of $0.5$ cm; further increase of the wire radius would not help. The field strength of $3$ T corresponds to the trap depth of $\sim 2$ mK $\sim 1.5\cdot 10^{-7}$ eV for $n$, the values typical for ultra-cold neutrons \cite{lusch69,stey69}. Further improvement can be obtained by a decrease of temperature of the wire to $1.5-1.7$ K or the use of superconductive wire with a larger ratio of NbTi/Cu. This may increase the trap depth by a factor of 2-3, thus making magnetic trapping of the full UCN energy range quite realistic.

Using the MGT for atoms with unpaired electron spin (H and D) in a high field seeking state should be much easier because of their much larger magnetic moments. We can reduce the current by three orders of magnitude, or, keeping the same current density, decrease the wire radius, or trap atoms at higher temperatures. Reducing the wire radius, one should take care of not violating the adiabaticity condition for the atomic motion. Keeping the same constraint of the fixed current density, the Larmor precession frequency scales as $\omega_L\sim r$, while the orbital rotation frequency scales as $\Omega\sim\ 1/\sqrt{r}$. However, even having a micrometer radius wire still does not violate adiabaticity, and such a trap could be realized for H and D.

Table \ref{Table1} presents typical parameters of various particles in the MGT.

\begin{table}
\centering
\begin{tabular}{|c|l|l|l|l|l|}
\hline
Particle & $I$ (A) & $r$ (cm) & $E$ (eV) & $\Omega/(2\pi)$ (Hz) & $v$ (m/s) \\
\hline
H & 10 & 1 & $5.8$ $10^{-9}$  & 16.5& 1.0 \\
H & 10 & 0.5 & $1.2$ $10^{-8}$  & 47.4& 1.5 \\
D & 10 & 1 & $5.8$ $10^{-9}$ & 11.8& 0.74 \\
D & 10 & 0.5 & $1.2$ $10^{-8}$ & 33.5& 1.1 \\
n & 10 000 &1 &$5.8$ $10^{-9}$ & 16.5& 1.0  \\
n & 10 000 &0.5 &$1.2$ $10^{-8}$ & 47.4& 1.5  \\
\hline
\end{tabular}
\caption{Typical parameters (orbit radius $r$, energy $E$, rotation frequency $\Omega$, velocity $v$) of a particle (hydrogen atom, deuterium atom, neutron) state bound in the magnetic field of linear current $I$.} \label{Table1}
\end{table}

\subsection{Gravitational quantum states}

The particle is confined vertically by the gravitational field and a mirror. This motion is quantized and described by GQSs. Such states were predicted \cite{lus78} and discovered \cite{nes02} for $n$, and predicted for $\bar{H}$ and $H$ atoms [32]. All details about the physical properties of such states can be found in the cited papers; here, we give only a summary of the main properties of these states in Table \ref{Table4} for the reader's convenience.

\begin{table}
 \centering
 \begin{tabular}{|c|l|l|l|l|}
  \hline
  $n$ & $\lambda_n$ &  $E_n$ (peV) & $E_n$ (Hz) & $z_n$ ($\mu m$)\\
  \hline
  1 & 2.338 & 1.407 &340.11 &13.726 \\
  2 & 4.088 & 2.461 &594.8  & 24.001\\
  3 & 5.521 & 3.324 &803.7  & 32.414\\
  4 & 6.787 & 4.086 &988.0  & 39.846\\
  5 & 7.944 & 4.782 &1156.3 & 46.639 \\
  6 & 9.023 & 5.431 &1313.2 & 52.974\\
  7 & 10.040& 6.044 &1461.4 & 58.945 \\
  \hline
 \end{tabular}
 \caption{Eigenvalues $\lambda_n$ (roots of Airy function), gravitational energies $E_n$ and classical turning points $z_n$ for neutrons, hydrogen and anti-hydrogen atoms in the Earth's gravitational field above a mirror.} \label{Table4}
 \end{table}

 The characteristic energy and spatial scales of such states are given by:
 \begin{equation}\label{lscale}
 \varepsilon_g=\sqrt[3]{ \hbar^2 m g^2/2},\mbox{ }
 l_g=\sqrt[3]{\hbar^2/(2m^{2}g)}.
 \end{equation}

\section{Coupling of vertical and radial motions}
Vertical and horizontal motions of particles are decoupled only to a finite precision. Below we consider phenomena which can mix them.

\subsection{The effect of wire non-verticality and mirror non-horizontality}

Although the precision of setting the mirror and wire directions can be high, the magnetic field would slightly deviate from eq. (\ref{Urad}), in particular due to the environment. An unavoidable vertical field gradient could result in false effects. To estimate these, we derive an expression for the magnetic field in the vicinity of a particle's circular trajectory.

\begin{eqnarray}
B=\frac{\mu_0 I}{2\pi \rho}, \rho=\sqrt{(r\cos(\varphi)\cos(\alpha)-z \sin(\alpha))^2+r^2\sin^2(\varphi)}.
\end{eqnarray}
Here $\rho$ is the distance between the wire and a particle, $\varphi$ is the particle angle in the plane, $z$ is the vertical coordinate of a particle above the mirror plane, $\alpha$ is the angle between the wire and vertical direction.

Below, we show that the ratio $z/r\ll 1$ is small for all states of interest. Taking into account the smallness of the deviation of $\alpha$ from the vertical direction, the potential energy is:
\begin{equation}
U(z)\simeq -\frac{\mu_0 \mu  I}{2\pi r} \left( 1+\frac{z \alpha \cos(\varphi)}{r}\right).
\end{equation}

The corresponding vertical  component of acceleration due to the wire non-verticality is:
\begin{equation}
a_m=a_0 \cos(\varphi), \mbox{ } a_0=\frac{\mu_0 \mu I}{2\pi m r^2}\alpha.
\end{equation}

Due to the periodicity of the function $\cos(\varphi(t))$, a gravitational energy correction due to the extra acceleration in the gradient magnetic field vanishes in the first order of the small parameter $a_0/g$:
\begin{equation}
\varepsilon_g ' = \frac{1}{2\pi}\int \left( \frac{m \hbar^2 \left(g+a_0 \cos(\varphi)\right)^2}{2}\right)^{1/3} d\varphi.
\end{equation}

The first non-vanishing correction to the unperturbed gravitation energy level $E_n=m g z_n$ appears in the second order:
\begin{equation}
\Delta E_n= -\frac{1}{18}\frac{m z_n a_0^2 }{g}.
\end{equation}

In Table \ref{Table2}, we present typical values of $a_0$ and the correction to the frequency shift between second and first GQSs for different orbits of a trapped particle.

\begin{table}
\centering
\begin{tabular}{|c|l|l|l|l|l|}
\hline
Particle & $I$ (A) & $\alpha$ & $R$ (cm) & $a_0$ (m/s$^2$) & $\Delta \omega_{21} $ (Hz) \\
\hline
H & 10 & $10^{-3}$ & 1& $0.055$&  $-0.003$    \\
H & 10 & $10^{-4}$ & 1& $0.0055$ & $-0.00003$   \\
H & 10 & $10^{-3}$ & 0.5& $0.22$ & $-0.05$  \\
n & 10000 & $10^{-3}$ & 1& $0.0055$ & $-0.003$     \\
n & 10000 & $10^{-4}$ & 1& $0.055$ &$-0.00003$   \\
n & 10000 & $10^{-3}$ & 0.5& $0.22$ &$-0.05$  \\
\hline
\end{tabular}
\caption{Extra acceleration and the gravitation transition frequency shift in the magnetic field of a linear current due to the non-verticality of wire alignment ($\Delta\omega_{21}=\Delta E_2- \Delta E_1$).} \label{Table2}
\end{table}

As one can see in Table \ref{Table1}, the particle rotation frequencies on orbits in the trap are small compared to the transition frequencies between low GQSs. Thus, no resonance effects could be found.

\subsection{Effect of a non-vanishing vertical gradient of the magnetic field}

A non-vanishing vertical gradient of the magnetic field could be due to a finite trap size. It might produce sizable effects which have to be compensated to a maximum degree. A residual gradient would result in a transition frequency shift proportional to the current. Thus, it could be extracted from experimental data by extrapolating the frequency shift to zero current.

\subsection{Effect of vibration of the wire and mirror}

The wire vibration effect can be estimated by assuming that the angle between the wire and the vertical direction is a periodic function of time, which is changing with an oscillation frequency%
\begin{equation}
\alpha(t)=\alpha_0\cos(\omega_v t).
\end{equation}

Then the perturbing potential is:
\begin{equation}
U(z,t)\simeq\frac{\mu_0 \mu  I}{2\pi r} \frac{z \alpha_0\cos(\omega_v t) \cos(\varphi)}{r}.
\end{equation}

Though its amplitude is small, as established above, oscillations of the perturbing potential (due to multiple modes of wire self oscillations) could appear to be in resonance with the transition frequency between GQSs as considered in \cite{cod12}. In this case, the transition probability $P_{ik}$ between initial ($i$) and final ($k$) GQSs is given by the Rabi-type expression [60]:
\begin{equation}
P_{ik}=\sin^2\left(\Omega_{ik} t\right).
\end{equation}
Here, the transition rate is:
\begin{equation}
\Omega_{ik}=\frac{\mu_0\mu I}{4\pi r}\frac{\alpha_0 l^3_g}{|z_i-z_k|^2}.
\end{equation}

The time $T_{ik}$ needed for the complete a transition from state $i$ to state $k$ is%
\begin{equation}
\label{WireOscillationTransitions}
T_{ik}=\frac{\pi}{2\Omega_{ik}}.
\end{equation}

Characteristic times $T_{12}$ are given in Table \ref{Table3}. 

\begin{table}
\centering
\begin{tabular}{|c|l|l|l|l|}
\hline
Particle & $I$ (A) & $\alpha_0$ & $R$ (cm) & $ T_{12}$ (s) \\
\hline
H & 10 & $10^{-3}$ & 1& $93.2$   \\
H & 10 & $10^{-4}$ & 1& $931.9$  \\
H & 10 & $10^{-3}$ & 0.5& $46.6$\\
n & 10000 & $10^{-3}$ & 1& $93.2$    \\
n & 10000 & $10^{-4}$ & 1& $931.9$  \\
n & 10000 & $10^{-3}$ & 0.5& $46.6$ \\
\hline
\end{tabular}
\caption{Transition time from first to second gravitational state due to wire oscillations in according with eq. (\ref{WireOscillationTransitions}).} \label{Table3}
\end{table}

\subsection{Effect of Earth's rotation}
The effect of Earth's rotation results in an additional Coriolis acceleration that a moving particle acquires in the non-inertial frame. The vertical component of such an acceleration $a_c$ of a particle trapped on a circular orbit is given by:
\begin{equation}
a_c= \Omega_E \cos(\Theta)\sqrt{\frac{\mu_0 \mu I}{\pi m r}}\cos(\varphi).
\end{equation}
Here, $\Omega_E=7.27\times 10^{-5}$ rad/s is the Earth's rotation frequency around its axis, $\Theta$ is a latitude of geographic position (45$^o$ in Grenoble). In case of an $H$ atom trapped in a circular orbit with radius $r=0.5$ cm and current $I=10$ A, the acceleration is $a_c=7.6 \times10^{-5}$ m/s$^2$.

After averaging over the trajectory, the first order Coriolis effect is canceled due to the periodic $cos(\varphi)$ factor. The second order Coriolis effect is well below the accuracy of our experiment and can be neglected.

\section{Feasibility of precision studies of GQS in the MGT}

\subsection{Loading/unloading the MGT}

The MGT can be loaded with $H$, $\bar{H}$, for instance, with rapidly switching the wire current. This is technically feasible due to the relatively small electric currents needed. For an atom velocity of $\sim 1$ m/s and a trap size of $\sim 10^{-2}$ m, the $\sim 10$ A current switching time should be significantly shorter than $\sim 10$ ms (see Table \ref{Table1}). The trap phase-space volume is estimated by assuming that all atoms with a velocity $v$ lower than the escape velocity $v_c$ for a given atom-wire distance would be captured. Then, the number of captured atoms is:
\begin{equation}
N=2f_0 \sqrt{\frac{\mu_0 \mu I}{\pi m}}\left(\sqrt{r_2}-\sqrt{r_1}\right).
\end{equation}
Here, $r_2$, $r_1$ are maximum and minimum radius of the trap, $f_0$ is an average density of atoms in a phase volume which is characterized by a maximum velocity $v_{max}=\sqrt{\mu_0 \mu I/(\pi m r_1)}$ and a spatial size $2 \pi (r_2^2-r_1^2)$.

In the case of $n$, a particularly elegant method of loading the MGT consists of producing UCNs directly in the trap filled in with superfluid $^4He$ \cite{golub75}. An attractive feature of this method is uniform occupation of the full phase-space volume available that is particularly important in view of the limited phase-space densities of UCNs.

Other methods include adiabatically changing the wire current or an additional uniform magnetic field in certain mirror geometries, spin-flip by a radio-frequency magnetic field, and various mechanical devices.

\subsection{Resonance Spectroscopy}
Methods of resonant gravitational spectroscopy have been developed in detail for $n$ and $\bar{H}$, for instance in \cite{pig14,vor14,bas15}, and can be easily extended to $H$. For $H$ in the MGT, one can observe the resonant changes in spatial density of the particles localized in GQSs above a mirror as a function of the oscillating frequency of an additional vertical magnetic field gradient. The resonant transitions result from the interaction of the magnetic dipoles of the trapped particles with the field gradient. The changes in spatial density are enhanced when the oscillating frequency coincides with the transition frequency $\omega_{nk}=(E_n-E_k)/\hbar$. The additional field is assumed to have the form:
\begin{equation}\label{Magn1}
\vec{B}(z,x,t)=B_0\vec{e}_z +\beta \cos(\omega t)  \left(z \vec{e}_z \right) .
\end{equation}
Here, $B_0$ is the amplitude of the static field component, $\beta$ is the oscillating magnetic field gradient. The oscillation frequency is determined by the transition frequency between the lowest GQSs; its typical value is $\omega\sim 10^3$ rad/s. For simplicity we omitted radial components of magnetic field, which do not influence the dynamics in the system. The static axial component $B_0$ provides a non-zero $z$ component of the atomic magnetic moment inducing the force in the oscillating field gradient. The strength of this component should not exceed the characteristic strength of the trapping field. Such a field configuration can be provided with a pair of coils in the anti-Helmholtz configuration arranged around the trap.

First, $H$ are prepared in the ground GQS. This is achieved by absorbing highly excited states using a scatterer/absorber plate lowered down to at a certain height $H_a$ above the mirror surface. Adjusting the $H_a$ value close to the characteristic delocalization height of the ground state $l_0\lambda_1<H_a$ (eq. \ref{lscale}) will effectively remove all other GQSs from the trap. This technique was used for in-beam spectroscopy of GQSs of $n$ \cite{nes05,mey14,pig14,vor14}. Vertical motion of the absorber over $\sim 10$ $\mu$m can be achieved with piezo-actuators.

At the second stage, the absorber is lifted up by $20-30$ $\mu$m to allow trapping a few exited GQSs. Oscillating vertical gradient of the magnetic field is applied at a frequency $\omega$. The field induces transitions from the ground to excited GQSs with the probability, which depends resonantly on the oscillating frequency $\omega_{1n} = (E_n-E_1)/\hbar$.

Third, the number of $H$ remaining in the ground GQS can be measured by placing absorber down again and eliminating $H$ in excited states. The numbers of $H$ in the ground state before the excitation and after can be measured by releasing them from the trap on to a detector.

The probability to excite $H$, initially prepared in the ground state, is given by the following expression:
\begin{equation}\label{Pdetect}
P(T)=|\sum_k C_k(T)\exp\left(-iE_k t/\hbar\right) \int_{H_d}^{\infty}\frac{\mathop{\rm Ai}(x-\lambda_k)}{\mathop{\rm Ai'}(-\lambda_k)} dx|^2.
\end{equation}
 Here, $T$ is the time of interaction of $H$ in the ground state with the oscillating magnetic field gradient. Assuming that the field frequency is close to the resonance $\omega_{1n}$ and the time is sufficiently large, $T\gg \hbar/\omega_{1n}$, one can get a simplified expression for the probability averaged over a period $\hbar/\omega_{1n}$:
\begin{equation}\label{PdetectN}
\bar{P}(T)=\exp\left(-\Gamma t/\hbar \right) \sum_n | C_n(T)|^2 \left | \int_{H_d}^{\infty}\frac{\mathop{\rm Ai}(x-\lambda_n)}{\mathop{\rm Ai'}(-\lambda_n)} dx\right |^2.
\end{equation}

The resonance frequency value corresponds to a maximum of loss of $H$ from the ground state as a function of the applied magnetic field frequency. Similarly, it is possible to measure transitions between any other pairs of low GQSs.

\subsection{Interferometry of quasi-stationary states}
Interferometry methods are based on the observation of a time distribution of detection events at a given $z$-location, or a vertical position distribution of detection events at a given time. An interference pattern can be observed if a pure initial state or a superposition of states is shaped.

Here, we analyze an example of the time evolution of an initially prepared wave-packet of $\bar{H}$ with a well-defined initial location $z_0$ above the mirror:
\begin{equation}
\Psi(z,t=0)=\sqrt{\frac{1}{\sigma \sqrt{\pi}}} \exp \left( -\frac{(z-z_0)^2}{2\sigma^2}\right),
\end{equation}
$\sigma$ is a spatial size of the initial state. 

We study vertical motion and ignore classical radial motion in the following.

Evolution of the wave-function $\Psi(z,t)$ is given by the following expression:
\begin{equation}
\Psi(z,t)=\int \Psi(z',0)K(z',z,t)dz'.
\end{equation}
Here $K(z,'z,t)$ is the particle propagator in GQSs:

\begin{equation}
K(z,'z,t)=\sum_i \frac{\mathop{\rm Ai}(z/l_g-\lambda_i)}{|\mathop{\rm Ai'}(-\lambda_i)|^2}\mathop{\rm Ai}(z/l_g-\lambda_i)\exp \left(-i \lambda_i \omega_g t \right).
\end{equation}
$\omega_g$ is a characteristic gravitational frequency:
\begin{equation}
\omega_g=\sqrt[3]{\frac{m g^2}{2\hbar}}=903.362 \mathop{ } s^{-1}.
\end{equation}
$\lambda_i$ is an eigenvalue (complex) of GQS.

In the limit $\sigma\ll l_g$, the following simplified expression for the wave function is valid:
\begin{equation}\label{PsiT}
\Psi(z,t)=\frac{\sqrt{2 \sigma \sqrt{\pi}}}{l_g} \sum_i \frac{\mathop{\rm Ai}(z_0/l_g-\lambda_i)}{|\mathop{\rm Ai'}(-\lambda_i)|^2}\mathop{\rm Ai}(z/l_g-\lambda_i)\exp \left(-i \lambda_i \omega_g t \right).
\end{equation}

Due to the weak annihilation of $\bar{H}$ on the surface, the mirror plays the role of detector. The corresponding rate of disappearance of $\bar{H}$ is given by expression \cite{vor11}:
\begin{equation}\label{Drate}
dP/dt=-\frac{2 \sigma \sqrt{\pi}}{l_g}\frac{\Gamma}{\hbar}\exp\left(-\Gamma t/\hbar\right) \sum_{i,j}\left( \frac{|\mathop{\rm Ai}(z_0/l_g-\lambda_i)|^2}{|\mathop{\rm Ai'}(-\lambda_i)|^2}\delta_{ij}+ 2(-1)^{i+j}\mathop{\rm Re}\frac{\mathop{\rm Ai}(z_0/l_g-\lambda_i)\mathop{\rm Ai}(z_0/l_g-\lambda_j)}{|\mathop{\rm Ai'}(-\lambda_i)\mathop{\rm Ai'}(-\lambda_j)|}\exp\left(-i \omega_{ij}t\right)\right).
\end{equation}
Here $\Gamma=\omega_g \mathop{\rm Im}\lambda_i$ is a quasi-stationary state width.

The  interference terms in expression (\ref{Drate}) are controlled by the frequencies $\omega_{ij}$ of transitions between GQSs. Measurements of the transition frequencies give access to the characteristic gravitational energy value $\varepsilon_g$ (\ref{lscale}).

\subsection{Measuring the momentum distribution in GQSs}

\subsubsection{Sudden mirror drop}
A vertical momentum (velocity) distribution $F(p)$ provides information about spatial properties of GQSs. A method to measure it for $\bar{H}$ can consist in a prompt downward shift of the mirror. The mirror acceleration $a$ should significantly exceed the free fall acceleration $g$. The vertical shift should significantly exceed the characteristic gravitational length scale $l_g$ (\ref{lscale}) and can be achieved using a piezo-actuator. Then, the initial superposition of GQSs starts falling freely at time $t_0$ down to the mirror (detector) installed at a distance $H_d$ below. One can use the sudden approximation to describe this motion. A fraction of the $\bar{H}$ annihilates; the remaining $\bar{H}$ are reflected from the surface of the mirror due to quantum reflection and continue bouncing until their full annihilation.

The initial momentum distribution $F(p)$ is mapped into the time distribution of free fall events \cite{crep19,nes19}:
\begin{equation}
\Phi\simeq mg|F\left(mg(t-t_f)\right)|^2.
\end{equation}
Here $\Phi$ is the flux of $\bar{H}$ falling on the annihilation surface, $t_f=\sqrt{2H_{d}/g}$ is the classical free fall time.

The momentum distribution $F(p,t_0)$ ($t_0$ is the moment of the sudden mirror drop) can be evaluated by Fourier transform of the distribution (\ref{PsiT}) taken at time $t_0$. By measuring the time distribution of free fall events, one obtains the momentum distribution of GQS, which gives access to the characteristic spatial scale of GQSs $l_g$ (\ref{lscale}).

\subsubsection{Prompt kick}
One could also use a prompt kick to make all $\bar{H}$ acquire the same upward momentum $p_0$. Such a kick could be achieved either by absorbing a photon from a laser beam, or by a prompt switching of a gradient magnetic field $W(t,z)=f(t-t_0)\mu z \partial B/\partial z $. $f(t-t_0)$ is a function, which characterizes the time dependence of gradient magnetic field and is localized around $t_0$ with a typical dispersion $\tau\ll 1/\omega_g$. In the limit $\tau\rightarrow 0$, the wave-function change after promptly switching the field is:
\begin{equation}
\Psi(z,t_0^+)=\exp \left(i p_0 z\right ) \Psi(z,t_0^-).
\end{equation}
Here
\[p_0=\mu  \partial B/\partial z \int f(t-t_0)dt. \]

The momentum distribution in an upstream detector is:
\begin{equation}
P\simeq mg|F\left(mg(t-t_f)+p_0 \right)|^2.
\end{equation}

Evaluating the energy and spatial gravitational scales from interferometry experiments and comparing these values with theory, one can conclude on a presence of extra interactions between atom and mirror at a micrometer scale.

\section{Conclusion}
We have proposed in this paper a new method for producing long confinement times in gravitational quantum states (GQSs) and a magneto-gravitational trap (MGT) for atoms, anti-atoms, neutrons and other particles possessing a magnetic moment. The Earth gravitational field and a reflecting mirror confine the particles in the vertical direction. The interaction of the particle's magnetic moment and the magnetic field originating from an electric current passing through a vertically installed wire confines particles in the radial direction, combined with the centrifugal acceleration of the particles. We underline that the observation time is the key parameter that defines the precision of measurements of such states. In case of anti-hydrogen atoms, one can achieve the observation time limited only by their annihilations in the surface [73]. In case of hydrogen atoms, storage times will be significantly longer due to a contribution from their specular reflection from the surface in the case of direct contact. In case of neutrons, storage times can approach the neutron lifetime. Our analysis shows that mixing of vertical and horizontal motions of particles can be controlled to an acceptable level not prohibiting precision measurements of GQSs in the MGT. We give examples, which prove feasibility of precision studies of GQS in the MGT.

These ideas can be applied in particular to the GRASIAN (co-authors of the present article), GBAR \cite{ind14} and GRANIT \cite{roul15} projects. S.V. thanks Academy of Findland for support (grant N.317141).

\section{References}

\end{document}